\documentclass[12pt]{article}
\usepackage{axodraw}
\begin{document}
\newcommand{\bqa}{\begin{eqnarray}}
\newcommand{\eqa}{\end{eqnarray}}
\newcommand{\nl}{\nonumber \\}
\newcommand{\ord}{{\cal O}}
\newcommand{\bfig}{\begin{center}\begin{picture}}
\newcommand{\efig}[1]{\end{picture}\\{\small #1}\end{center}}
\newcommand{\flin}[2]{\ArrowLine(#1)(#2)}
\newcommand{\wlin}[2]{\DashLine(#1)(#2){2.5}}
\newcommand{\zlin}[2]{\DashLine(#1)(#2){5}}
\newcommand{\glin}[3]{\Photon(#1)(#2){2}{#3}}
\newcommand{\lin}[2]{\Line(#1)(#2)}
\newcommand{\sof}{\SetOffset}
 \begin{flushright}
{\large UG-FT-162/04}

{\large CAFPE-32/04}

{\large \today}
\end{flushright}

\vspace{0.3cm}

\begin{center}
{\Large\bf Recursive numerical calculus of one-loop tensor integrals}

\vspace{1cm}

{\Large F. del Aguila and R. Pittau}
\footnote{On leave of absence from Dipartimento di 
Fisica Teorica, Torino and INFN Sezione di Torino, Italy.}

\vspace{1cm}
{\em Departamento de F{\'\i}sica Te\'orica y del Cosmos 

and Centro Andaluz de 
F{\'\i}sica de Part{\'\i}culas Elementales (CAFPE), 

Universidad de Granada, E-18071 Granada, Spain}
\end{center}

\vspace{1cm}
 
\begin{abstract}
\noindent\small 
A numerical approach to compute tensor integrals in one-loop calculations
is presented. The algorithm is based on a recursion relation 
which allows to express high rank tensor integrals as a function of 
lower rank ones.
At each level of iteration only inverse square roots of 
Gram determinants appear. 
For the phase-space regions where Gram determinants are so small
that numerical problems are expected, we give
general prescriptions on how to construct reliable approximations
to the exact result without performing Taylor expansions.
Working in $4+\epsilon$ dimensions does not require an analytic 
separation of ultraviolet and infrared/collinear divergences, and,
apart from trivial integrals that we compute explicitly, no
additional ones besides the standard set of scalar 
one-loop integrals are needed.
\end{abstract}

\vspace*{3.8truecm}


\vspace*{0.2truecm}


\thispagestyle{empty}
\clearpage
\setcounter{page}{1}

\section{Introduction}
\label{sec:intro}
Computing radiative corrections in high-energy Particle Physics demands an 
increasing capability of manipulating complicated objects. 
When the number of particles undergoing the 
scattering process grows, formidable complications arise.
For example, to date no complete QCD or Electroweak (EW) calculation 
exists at the one-loop level 
describing processes involving 2 ingoing and 4 outgoing particles.
On the other hand, multi-particle processes need to be studied
with great accuracy at the next generation of $pp$ and $e^+ e^-$ colliders, 
hence we cannot escape the technical subject of efficiently 
compute radiative corrections, at least at the one-loop level.
To do this three main problems must be faced:
the large number of Feynman diagrams, 
the reduction of tensor integrals to scalar ones, and 
the control over the numerical inaccuracies. 

As for the first issue, while several tree level algorithms 
to compute amplitudes without
making explicit reference to Feynman diagrams exist
\cite{LO}, so far there is no equivalent working technique 
when loops enter the game.
At any rate, for moderate values of external particles
the number of contributing Feynman diagrams can be in principle manageable.
For example, this number is of the order of thousand for the EW process 
$e^+ e^- \to \mu^- \bar \nu_\mu u \bar d$, 
depending on the chosen gauge.
What really renders the calculation difficult is the fact that
each diagram still requires a lot of work to be computed.
This is the second issue mentioned above.

In order to be concrete, let us sketch out a typical 
one-loop calculation performed in $n= 4+\epsilon$ dimensions.
The corresponding amplitude $A^{(n)}$ can be written as 
a sum of tensor integrals $I^{(n)}$ times external tensors $S$.
\bqa
\label{eq:16}
A^{(n)} = \sum_{m,j,i} I^{(n)}_{m,j;\,\mu_1 \cdots \mu_i}
\,\,S_{m,j}^{\mu_1 \cdots \mu_i}\,,
\eqa
where
\bqa
\label{eq:2}
 I^{(n)}_{m,j;\,\mu_1 \cdots \mu_i}
= 
\int d^n q 
\frac{q_{\mu_1} \cdots q_{\mu_i}}{D_0 D_1 \cdots D_m}\,,
~~D_k= (q+p_k)^2-m^2_k\,,~~k=0,\cdots ,m\,,~~p^\mu_0= 0 \,,
\eqa
with $j$ labelling the different momenta $p_k$ entering and 
masses $m_k$ running the loop,
and where $S_{m,j}^{\mu_1 \cdots \mu_i}$ only depend on the 
external kinematics.
The $(m+1)$-point tensor one-loop integrals 
are usually evaluated in two steps.
First one writes down the most general decomposition in terms of 
the external momenta and the metric tensor
\bqa
\label{decomp}
I^{(n)}_{m,j;\,\mu_1 \cdots \mu_i}
= \!\!\!\!\!\!\!\!\!\!\!
\sum_{
\begin{tabular}{l}
$s,s_1, \cdots, s_m$ \\
$2s+\sum s_k= i$ 
\end{tabular}
}\!\!\!\!\!\!\!\!\!\!
c^{(n)}_j\,(s,s_1, \cdots, s_m)\,
\left\{
[g]^s [p_1]^{s_1} \cdots [p_m]^{s_m}
\right\}_{\mu_1 \cdots \mu_i}\,, 
\eqa
where
$\left\{
[g]^s [p_1]^{s_1} \cdots [p_m]^{s_m}
\right\}_{\mu_1 \cdots \mu_i}$ corresponds to the 
symmetrical tensor combination, each term of which is 
constructed from $s$ metric tensors, $s_1$ momenta 
$p_1$, $\cdots$, and $s_m$ momenta $p_m$. For example, 
\bqa
\left\{g p_1\right\}_{\mu_1 \mu_2 \mu_3}= 
 g_{\mu_1\mu_2} {p_1}_{\mu_3}
+g_{\mu_1\mu_3} {p_1}_{\mu_2}
+g_{\mu_2\mu_3} {p_1}_{\mu_1}\,.
\eqa  
Secondly one looks for explicit expressions 
for the scalar coefficients 
$c^{(n)}_j\,(s,s_1, \cdots, s_m).$
In the Passarino-Veltman treatment 
\cite{Passarino:1978jh} they are expressed in terms 
of a minimal set of scalar one-loop integrals 
with denominators raised to the first 
power only \cite{'tHooft:1978xw}
\bqa
\label{minimal}
 I^{(n)}_{m,j}
= \int d^n q 
\frac{1}{D_0 D_1 \cdots D_m}\,.
\eqa
In other methods \cite{Tarasov:1996br,Davydychev:1991va}
the decomposition is performed in terms of an enlarged set of
scalar integrals with shifted dimensionality and denominators 
raised to generic powers
\bqa
\label{shifted}
 I^{(n+\nu)}_{m,j,\nu_j}
= \int d^{(n+\nu)} q 
\frac{1}{D_0^{\nu_0} D_1^{\nu_1} \cdots D_m^{\nu_m}}\,.
\eqa
Of course, there are relations among the integrals in 
Eq. (\ref{shifted}) and the  minimal set in
Eq. (\ref{minimal}), which have been extensively studied
in Ref. \cite{Tarasov:1996br}.
For large $m$ values
the algebraical complexity of the described methods
quickly becomes overwhelming, 
mainly due to the rapidly increasing number of kinematic 
variables present in the problem.
For example, the application of Eq. (\ref{decomp}) to
$I^{(n)}_{4,j;\,\mu_1 \mu_2 \mu_3}$ generates 24  
scalar coefficients.

An alternative to this procedure is a numerical approach.
The ideal situation would be that one simply 
writes down Eq. (\ref{eq:16}) while all the rest is handled 
by a numerical program.
The main problem with such an approach is handling ultraviolet and 
infrared/collinear singularities.
Along this road a complete formalism has been recently 
presented in the framework of QCD by Giele and Glover \cite{Giele:2004iy}.
In their method one first analytically separates 
the divergent contributions arising
from the tensor integrals in Eq. (\ref{eq:2}), using
the techniques of Refs. 
\cite{Binoth:1999sp,Duplancic:2003tv,Binoth:2002xh}. 
Then one computes numerically the kinematical coefficients 
of the resulting finite 4-dimensional integrals.
A different approach is sewing tree amplitudes together
to construct loop amplitudes, as proposed in Ref.
\cite{Bern:1994cg}. Another way is combining virtual 
and real contributions to cancel the divergences 
in the loop integration
\cite{Soper:1999xk}, or constructing counter-terms 
diagram by diagram \cite{Nagy:2003qn}.
Finally, the authors of Ref. \cite{Ferroglia:2002mz}
develop a pure numerical approach
where in the Feynman parameter space
any one-loop integrand is cast in a form well suited 
for numerical calculation, in such a way that
all possible divergences get automatically extracted.   

In this paper we present a new method in which almost
all the work can be performed numerically: 
the tensor integrals in Eq. (\ref{eq:2}) are
numerically reduced to the minimal set in Eq. (\ref{minimal}), 
and the ultraviolet and infrared/collinear divergences
are controlled without performing any explicit subtraction.
Then, any amplitude can be calculated simply contracting
the numerically computed tensor integrals with the external 
tensors $S_{m,j}$ (see Eq. (\ref{eq:16})).
Our main result will be the derivation of the set of recursion 
relations that link high rank tensor integrals to lower rank 
ones, and which allow the numerical reduction of the former 
to the minimal set in Eq. (\ref{minimal}).
On the other hand, 
our solution to the problem of handling divergences is 
splitting beforehand any tensor into its pure 4-dimensional part plus
any other additional contributions, which are trivial to compute. 
Such a procedure renders unnecessary an analytical 
separation of the divergent parts, as required by the recursion
relations of Ref. \cite{Binoth:1999sp}, and minimizes the 
analytic work.
At the end all divergences are contained in the pole parts
of the scalar integrals, parts to which
one can give any value to numerically check all relevant cancellations.
Furthermore, internal and/or external masses do not 
pose any particular problem, so that the method can be applied 
to both QCD and EW calculations.

The tensor integrals are usually well behaved
when two or more momenta become linearly dependent, as observed in Refs.
\cite{Devaraj:1997es, Stuart:1989de, Stuart:1987tt}.
However, in general reduction formulae introduce
inverse powers of Gram determinants 
\bqa
\label{Gram}
\Delta _{1\,\cdots\, m} =
{\rm Det}\, (p_i \cdot p_j)\,,~~~~{1 \le i,j \le m}\,
\eqa
in the decomposition terms. So when they go to zero, large 
cancellations among the different terms must take place, 
giving rise to numerical instabilities. This is the third 
problem one has to face.
In Ref. \cite{Campbell:1996zw} 
this is solved by systematically 
building up combinations of well behaved
functions in the limit of vanishing Gram determinants.
The drawback of this approach is that the class of needed 
loop functions has to be enlarged, including integrals 
computed in higher dimensions, such as those in Eq. (\ref{shifted}). 
When things are re-expressed in terms of (4+$\epsilon$)-dimensional 
integrals, Gram determinants are reintroduced
and explicit Taylor expansions are required
to deal with the problematic phase-space regions.
A possible solution 
is presented in Ref. \cite{Beenakker:2001rj},
where an extrapolation from the inner phase-space region
is used for the {\em dangerous} points.
In the numerical approach of Ref. \cite{Ferroglia:2002mz}
all scalar coefficients $c^{(n)}$ in Eq. (\ref{decomp}) are
cast in a form well suited for numerical evaluation.
In Ref. \cite{Denner:2002ii}
$4$-dimensional pentagon like tensors are reduced to box like tensors, 
avoiding the occurrence of rank $m=4$ Gram determinants. 
Finally, the case of {\em exactly} zero Gram determinant is solved 
in Ref. \cite{Binoth:1999sp} using the pseudo-inverse of the Gram matrix.

In our method part of the Gram determinant singularities
compensates in such a way that at each step of the iteration only 
inverse square roots of Gram determinants appear, improving the 
numerical stability of the calculation.
Moreover, the expressions can be very naturally arranged in groups 
which are well behaved in the limit of linearly dependent external momenta,
allowing to keep as local as possible the numerical cancellations 
that occur among the loop functions.
The case of {\em exactly} zero Gram determinants can be treated without
any problem, and for the problematic regions of nearly vanishing 
Gram determinants, we show how to systematically construct 
reliable approximations to the exact result.  
Unlike in Ref. \cite{Campbell:1996zw} building up such approximations 
does not require explicit Taylor expansions.

The paper is organised as follows.
Our master formula for the general 4-dimensional case is presented 
in next Section.
The algorithm is extended to the $n$-dimensional case 
in Section \ref{sec:2}.
The particular cases of 3-point and rank one tensors,
to which our general formula does not apply, are discussed 
in Sections \ref{sec:cfunctions} and \ref{sec:rankone}, 
respectively. Whereas in Section \ref{sec:coll} 
we perform a detailed study of the dangerous collinear 
and coplanar configurations.
Section \ref{sec:Summary} is devoted to conclusions; and
technical details are worked out in three Appendices.

\section{The 4-dimensional method}
\label{sec:1}
In the following we shall show our master formula to reduce 
high rank tensor integrals to lower rank ones. 
Let us first clarify the notation. 
Throughout the paper we drop the index $(n)$ from loop 
integrals when the space-time is 4-dimensional. We also omit the index 
$j$ for simplicity, although it is understood that an
$(m+1)$-point integral depends on $m$ external momenta and 
$m+1$ internal masses.
$m$-point tensor integrals obtained from
an $(m+1)$-point one by eliminating the $k^{th}$ denominator 
are written specifying the index of the dropped denominator 
as an argument
\bqa
\label{eq:1a}
 I_{m-1;\,\mu \cdots \nu}\,(k)= 
\int d^4 q \frac{q_\mu \cdots q_\nu}{D_0 D_1 
\cdots D_{k-1} D_{k+1} \cdots D_m}\,;
\eqa
and as usual tensor indices are raised and lowered with the metric tensor.

Our main result is the recursion relation for reducing 4-dimensional 
tensor loop integrals with $m\,>\,2$ and rank higher than 1,
\bqa
\label{eq:15}
I_{m;\,\mu \nu \rho \cdots \tau}  &=& \frac{\beta}{2 \gamma}\, 
T_{\mu  \nu \lambda \sigma}\,\left\{J^{\lambda\sigma}_{m;\,\rho \cdots \tau}
\right\}
~-~\frac{1}{4 \,\gamma} T_{\mu \nu} 
\left\{m_0^2  I_{m;\,\rho \cdots \tau} 
+I_{m-1;\,\rho \cdots \tau}(0)
\right\}  \nl
&-&\frac{1}{4 \,\gamma}
T_{\mu \nu \lambda} 
\left\{f_{30} I^{\lambda}_{m;\,\rho \cdots \tau}      
             +I^{\lambda}_{m-1;\,\rho \cdots \tau}(3)
             -I^{\lambda}_{m-1;\,\rho \cdots \tau}(0)
 -\frac{2 \beta}{\gamma} p_{3\alpha}\,J^{\alpha\lambda}_{m;\,\rho \cdots \tau}
 \right\}\,,
\eqa
where
\bqa
\label{eq:j}
J^{\lambda\sigma}_{m;\,\rho \cdots \tau} &=&
(f_{10} r_2^\lambda + f_{20} r_1^\lambda) 
I^{\sigma}_{m;\,\rho \cdots \tau}
+r_2^\lambda I^{\sigma}_{m-1;\,\rho \cdots \tau}(1)
+r_1^\lambda I^{\sigma}_{m-1;\,\rho \cdots \tau}(2) \nl
&-&(r_1^\lambda+r_2^\lambda) I^{\sigma}_{m-1;\,\rho \cdots \tau}(0)\,,
\eqa
and $T_{\mu  \nu \lambda \sigma}$, $T_{\mu  \nu \lambda}$,
$T_{\mu  \nu}$ and $r^\lambda_{1,2}$ only depend on 
three linearly independent external momenta, which we assume 
to be $p_{1,2,3}\,$. The scalar factors $\beta$ and 
$\gamma$ are only functions of $p_{1,2}\,$; whereas 
\bqa
\label{eq:9}
f_{k0} &=& m_k^2-m_0^2-p_k^2  \,.
\eqa
Eq. (\ref{eq:15}) can then be iterated to
compute numerically all 4-dimensional tensors $I_{m,j;\,\mu_1 \cdots \mu_i}$,  
without explicit tensor decomposition, starting from the standard
set of scalar loop functions in Eq. (\ref{minimal}). 
Notice that the variable shift $q \to q-p_1$ is needed 
before applying the next iteration to any tensor integral having $(0)$ 
as an argument. 

In order to derive Eq. (\ref{eq:15}), we need to express 
the product of two loop momenta as a sum of terms with at 
most one loop momentum times loop denominators, or internal 
masses, and external tensors (Eq. (\ref{eq:13}) below). 
First, we write the internal momentum as a sum of  
four judiciously chosen massless vectors
\bqa
\label{eq:6}
q = \sum_{i=1}^{4}\,c_i \ell_i\,,~~~~\ell_i^2=0\,. 
\eqa
Following Ref. \cite{Pittau:1997mv} we construct 
$\ell_{1,2}$ from two independent external momenta, which we 
assume to be $p_{1,2}\,$, 
\bqa
\label{eq:3}
p_1 = \ell_1+\alpha_1 \ell_2\,,~~~p_2 =  \ell_2+\alpha_2 \ell_1\,.
\eqa
Then
\footnote{\label{foot}
When one or both $p_i^2$ vanishes, $\beta= 1$, and
\bqa
\label{eq:5}
\begin{tabular}{lllll}
$p_1^2 =   0$, & $p_2^2 \ne 0$  &$\Rightarrow$ 
            & $\alpha_1 = 0$, &$\alpha_2 = {p_2^2}/(2\,(p_1 \cdot p_2))$\,, \\
$p_1^2 \ne 0$, & $ p_2^2 =  0$  &$\Rightarrow$  
            & $\alpha_1 = p_1^2/(2\,(p_1 \cdot p_2))$,  &$\alpha_2 = 0$\,, \\
$p_1^2 =   0$, & $ p_2^2  =  0$ &$\Rightarrow$
            & $\alpha_1 = 0$, &$\alpha_2 = 0$\,.
\end{tabular}
\nonumber
\eqa
These limits are smoothly approached taking for $\alpha_1$
the solution with $-\sqrt{\Delta}$ 
($+\sqrt{\Delta}$) when $(p_1 \cdot p_2)> 0$ ($(p_1 \cdot p_2)<0$), 
which is what we stand for $\mp \sqrt{\Delta}$.}
\bqa
\label{eq:4}
\begin{tabular}{ll}
$\ell_1 = \beta\,(p_1-\alpha_1\, p_2)$, & 
$\ell_2 = \beta\,(p_2-\alpha_2\, p_1)$, \\\\ 
$\alpha_1 =  ((p_1 \cdot p_2) \mp \sqrt{\Delta})/p_2^2$, & 
$\Delta \equiv - \Delta _{12}=  (p_1 \cdot p_2)^2-p_1^2 p_2^2$, \\\\ 
$\alpha_2=  \alpha_1\,p_2^2/p_1^2$,  & 
$\beta = {1}/(1-\alpha_1 \alpha_2)$. 
\end{tabular}
\eqa
Note that $\ell_{1,2}$ can be also complex.
The other two independent massless vectors $\ell_{3,4}$ are taken to be
\bqa
\label{eq:l34}
\ell_3^\mu = \bar v(\ell_1)\, \gamma^\mu\,       \omega^-\, u(\ell_2)\,,~~~
\ell_4^\mu = \bar v(\ell_2)\, \gamma^\mu\,       \omega^-\, u(\ell_1)\,,~~~
\omega^{-}  = (1- \gamma_5)/2\,,
\eqa
thus fulfilling the equalities (see Appendix \ref{app:1})
\bqa
\label{eq:5a}
(\ell_{3,4} \cdot \ell_{1,2}) =    
\ell^2_3 = \ell^2_4 = 0\,,~~~(\ell_3 \cdot \ell_4) = 
-4 (\ell_1 \cdot \ell_2)\,.
\eqa
Using them the coefficients $c_i$ in Eq. (\ref{eq:6}) can be simply written
\bqa
c_1 = \frac{(q\cdot\ell_2)}{(\ell_1\cdot\ell_2)}\,,~~~
c_2 = \frac{(q\cdot\ell_1)}{(\ell_1\cdot\ell_2)}\,,~~~
c_3 =-\frac{(q\cdot\ell_4)}{4\,(\ell_1\cdot\ell_2)}\,,~~~
c_4 =-\frac{(q\cdot\ell_3)}{4\,(\ell_1\cdot\ell_2)}\,.
\eqa
Now, it is convenient to distinguish between 
$\ell_{1,2}$ and $\ell_{3,4}$ in Eq. (\ref{eq:6}) 
for the contribution of the first two 4-vectors can be expressed 
as a sum of products of denominators, internal masses and external momenta, 
\bqa
\label{eq:8}
 q_{\mu} &=& \frac{\beta}{\gamma} D_\mu-\frac{1}{2\,\gamma} Q_\mu \,,
\nl
 D_{\mu} &=& \frac{1}{\beta}\,
[
2\,(q\cdot\ell_1) \ell_{2\mu}+
2\,(q\cdot\ell_2) \ell_{1\mu}
]
\nl
&=& \left\{ 
                 f_{10}\,r_{2\mu}+f_{20}\,r_{1\mu} 
                 +D_1\,r_{2\mu}+D_2\,r_{1\mu}-D_0\,(r_{1\mu} + r_{2\mu})
          \right\} \,,
\nl
 Q_{\mu} &=& (q \cdot \ell_3)\,\ell_{4\mu}+(q \cdot \ell_4)\,\ell_{3\mu}\,,~~~
\gamma~=~2 \frac{(p_1 \cdot p_2)}{1+\alpha_1\,\alpha_2} 
~=~ 2 \,(\ell_1 \cdot \ell_2)\,,
\eqa
where we have made use of
\bqa
\label{eq:9a}
(q \cdot p_k) = \frac{1}{2} \left[ D_k-D_0 + f_{k0} \right]\,,
\eqa
with $f_{k0}$ defined in Eq. (\ref{eq:9}), and
\bqa
\begin{tabular}{ll}
$r_1 = (\ell_1-\alpha_1\, \ell_2)$,\,\, &  $r_2 = (\ell_2-\alpha_2\, \ell_1)$.
\end{tabular}
\eqa
Then applying Eq. (\ref{eq:8}) twice and symmetrizing
on $\mu \nu$, we find
\bqa
\label{eq:10}
q_\mu q_\nu &=& \frac{\beta}{2 \gamma} \left[D^\lambda q^\sigma \right]
\left\{
 g_{\lambda \mu} \left(g_{\sigma \nu}-\frac{t_{\sigma \nu}}{2\gamma} \right) 
+g_{\lambda \nu} \left(g_{\sigma \mu}-\frac{t_{\sigma \mu}}{2\gamma} \right) 
\right\} 
+\frac{Q_\mu Q_\nu}{4 \,\gamma^2}\,,  \nl
t_{\rho \tau} &=& \ell_{3\rho}\ell_{4\tau}
                 +\ell_{4\rho}\ell_{3\tau}\,.
\eqa
The factor in square brackets contains reconstructed denominators while
the factor in curly brackets only depends on the external kinematics.
The $Q_{\mu} Q_{\nu}$ term can be also decomposed
using the properties of the 4-vectors $\ell_{3,4}$ 
(see Appendix \ref{app:1} for details)
%
\bqa
\label{eq:11}
(q \cdot \ell_3)(q \cdot \ell_4) &=& 4\,(q \cdot \ell_1)
\,(q \cdot \ell_2) -2\,q^2\,(\ell_1 \cdot \ell_2)~\equiv~
\beta q^\alpha D_\alpha -\gamma q^2\,,\nl
(q \cdot \ell_{3(4)})(q \cdot \ell_{3(4)}) &=& \frac{2}{(b \cdot \ell_{4(3)})}
\left\{ [q^2 (\ell_1 \cdot \ell_2)-2
           (q \cdot \ell_1)(q \cdot \ell_2)]\,
           (b \cdot \ell_{3(4)}) \right. \nl
    &+&\left.2\, 
         [(q \cdot \ell_1)(b \cdot \ell_2)
         -(q \cdot b)(\ell_1 \cdot \ell_2)
         +(q \cdot \ell_2)(\ell_1 \cdot b)]\,
          (q \cdot \ell_{3(4)}) 
            \right\} \nl
&\equiv&
\frac{1}{(b \cdot \ell_{4(3)})}
\left\{ [\gamma q^2-\beta q^\alpha D_\alpha]\,
           (b \cdot \ell_{3(4)}) \right. \nl
    &-&\left.\, 
         [\gamma\,(q\cdot b)-\beta b^\alpha D_\alpha]\,
          (q \cdot \ell_{3(4)}) 
            \right\}\,,~~~\forall\, b\ne \ell_{1,2}\,. 
\eqa
Now, choosing the arbitrary 4-vector $b= p_3$ and 
reconstructing again denominators, we get
\bqa
\label{eq:12}
Q_{\mu}Q_{\nu} &=& \beta \left[D^\rho q^\alpha \right]
\left\{ 
g_{\rho \alpha} T_{\mu \nu} \right\}  
- \gamma\, [D_0+m_0^2] \left\{T_{\mu \nu} \right\} \nl
&+& \left[ 2\, \beta \, p_{3\alpha}\, [D^\alpha q^\lambda]
- \gamma\, (D_3-D_0+f_{30}) q^{\lambda}\right] 
\left\{T_{\mu \nu \lambda}  \right\},
\nl
T_{\mu \nu \lambda} &=& 
 \frac{\ell_{3\mu}\,\ell_{3\nu}\,\ell_{4\lambda}}{(\ell_3 \cdot p_3)}  
+\frac{\ell_{4\mu}\,\ell_{4\nu}\,\ell_{3\lambda}}{(\ell_4 \cdot p_3)},  \nl
T_{\mu \nu} &=& t_{\mu\nu}-p_3^{\sigma} T_{\mu \nu \sigma}\,.
\eqa
Finally, inserting Eq. (\ref{eq:12}) into  Eq. (\ref{eq:10}) we obtain 
the decomposition relation for the product of two loop momenta
\bqa
\label{eq:13}
q_\mu q_\nu &=& \frac{\beta}{2 \gamma} \left[D^\lambda q^\sigma \right] 
T_{ \mu \nu \lambda \sigma }
-\frac{1}{4 \,\gamma}\left[D_0+m_0^2 \right] T_{\mu \nu} \nl
&-&\frac{1}{4 \,\gamma}
\left[\left((D_3-D_0+f_{30}) -\frac{2 \beta}{\gamma} p_{3\alpha}\,D^\alpha
\right) q^\lambda
\right] T_{\mu \nu \lambda}\,,
\eqa
where
\bqa
\label{eq:14}
T_{\mu \nu \lambda \sigma} &=&
 g_{\mu \lambda} 
 \left(g_{\nu \sigma}-\frac{t_{\nu \sigma}}{2 \gamma} \right )
+g_{\nu \lambda} 
 \left(g_{\mu \sigma}-\frac{t_{\mu \sigma}}{2 \gamma} \right )
+\frac{g_{\lambda \sigma}}{2 \gamma}\, T_{\mu \nu}\,.
\eqa
Then Eq. (\ref{eq:15}) follows after dividing Eq. (\ref{eq:13}) by the
$(m+1)$ denominators in Eq. (\ref{eq:2}), 
multiplying by the remaining 4-vectors $q_\rho \cdots q_\tau$
and integrating over $d^4 q$. 
As it stands, Eq. (\ref{eq:15}) is valid in non exceptional
phase-space regions where $p_{1,2,3}$ are all independent, 
otherwise zeros occur in the denominators.
In Section \ref{sec:coll}, we shall consider
the case of exceptional or nearly exceptional configurations.
The reason why it is convenient to group terms 
in the way we have done will be clear there.

As a last remark we observe that the rank of the tensor integral on 
the l.h.s. of Eq. (\ref{eq:15}) 
should be at least 2 and $m \ge 3$, otherwise there are  
not enough external momenta to perform the decomposition neither 
denominators to cancel the reconstructed ones. 
These two particular cases are discussed in Sections \ref{sec:rankone} 
and \ref{sec:cfunctions}, respectively. 

\section{The method in $n$ dimensions}
\label{sec:2}
The derivation of Eq. (\ref{eq:15}) breaks down when working 
in $n$ dimensions because Eq. (\ref{eq:6}) is only valid in
4 dimensions.
To deal with the $n$-dimensional case we need
to reduce the problem to 4-dimensional tensors,
that we know how to handle. As in Refs.
\cite{Pittau:1997mv,Bern:1995db,Bern:1991aq,Veltman:1988au} 
only unobservable objects are considered to live
in $n= 4+\epsilon$ dimensions, with $4$ and $\epsilon$-dimensional
quantities always orthogonal to each other. 
In particular, only the integration momentum $q$  is  $n$-dimensional
in our integrals.
For notational purposes from now on we put a bar over 
$n$-dimensional quantities and a tilde 
over $\epsilon$-dimensional objects. For example,
\bqa
\label{eq:18}
\bar q = q + \tilde q\,,~~~~{\bar q}^2= q^2+{\tilde q}^2\,,
\eqa
where $q$ is purely 4-dimensional. 
Being more explicit, the rank $i$ $n$-dimensional tensor integrals
we want to evaluate read
\bqa
\label{eq:19}
I^{(n)}_{m;\,\bar\mu_1 \cdots \bar \mu_i}
= \int d^n\bar q \frac{1}{\bar D_0 \cdots\bar  D_m}
\prod_{k= 1}^i \bar q_{\mu_k}
\,,~~~~
\bar D_k= (\bar q+p_k)^2-m^2_k\,.
\eqa
Using Eq. (\ref{eq:18}) to split the numerator momenta 
of Eq. (\ref{eq:19}), we get
\bqa
\label{eq:20}
\prod_{k= 1}^i \bar q_{\mu_k} &=& 
  \prod_{k= 1}^i q_{\mu_k}   
 +\sum_{s_1=1}^i \tilde q_{\mu_{s_1}}\prod_{k \ne s_1}^i q_{\mu_k}   \nl
&+&\sum_{s_1>s_2}^i \tilde q_{\mu_{s_1}} \tilde q_{\mu_{s_2}}
    \prod_{k \ne s_1,s_2}^i q_{\mu_k}  
 + \cdots + \prod_{k= 1}^i \tilde q_{\mu_k}\,.  
\eqa
Since the momenta $p_k$ are purely 4-dimensional,
all terms containing an odd number of $\tilde q$ in the numerator 
vanish, and all terms with an even number of $\tilde q$ 
can be only proportional to symmetric combinations of the metric tensor 
in $\epsilon$ dimensions
\bqa
\label{eq:21}
\tilde g_{\mu \nu} \equiv g_{\tilde \mu \tilde \nu}\,,~~
\tilde g_{\mu \nu \rho \sigma} \equiv 
 \tilde g_{\mu \nu} \,\tilde g_{\rho \sigma}
+\tilde g_{\mu \rho} \,\tilde g_{\nu \sigma}
+\tilde g_{\mu \sigma} \,\tilde g_{\nu \rho}\,,~~{\rm etc.}
\eqa
Therefore
\bqa
\label{eq:22}
I^{(n)}_{m;\,\mu_1 \cdots \mu_h \tilde \nu_1 \cdots \tilde \nu_{2\ell +1}}
\equiv
\int d^n\bar q \frac{1}{\bar D_0 \cdots\bar  D_m}
\prod_{k= 1}^h  q_{\mu_k}
\prod_{s= 1}^{2\ell +1} \tilde q_{\nu_s} = 0\,,
\eqa
and
\bqa
\label{eq:23}
I^{(n)}_{m;\,\mu_1 \cdots \mu_h \tilde \nu_1 \cdots \tilde \nu_{2\ell}} =
\Gamma\left(\frac{\epsilon}{2}\right)\,
\frac{\tilde g_{\nu_1 \cdots \nu_{2\ell}}}{2^\ell\, 
\Gamma\left(\frac{\epsilon}{2}+\ell\right)}
 \!\int d^n\bar q \frac{{\tilde q}^{2\ell}}{\bar D_0 \cdots\bar  D_m}
\prod_{k= 1}^h  q_{\mu_k}\,,
\eqa
where ${\tilde q}^2= \tilde q_\mu \tilde q^\mu$ and $\ell= 0,1,2,\cdots$.
As it is clear from the previous equation, a new class of 
integrals involving ${\tilde q}^{2 \ell}$ appears.
We will use for these objects the notation $I^{(n;\,2 \ell)}$. 
For example, the last integral in Eq. (\ref{eq:23}) is denoted by
$I^{(n;\,2 \ell)}_{m;\,\mu_1 \cdots \mu_h}$.
Such integrals are very easy to evaluate at ${\cal O}(1)$, and we 
do it in Appendix \ref{appa}.
One can argue that, since a pole in ${1}/{\epsilon}$ appears
in Eq. (\ref{eq:23}), higher order terms might be needed.
However, the $\epsilon$-dimensional tensor indices
in Eq. (\ref{eq:23}) can only be contracted with $\epsilon$-dimensional
indices (typically combinations of $\tilde g_{\mu\nu}$ tensors), otherwise
they give zero because of the orthogonality between 4-dimensional
and $\epsilon$-dimensional spaces.
In this contraction the pole cancels out, so that 
${\cal O}(\epsilon)$ terms can be safely neglected in 
the physical limit $\epsilon \to 0$.

Finally, the purely 4-dimensional integral 
$I^{(n)}_{m;\,\mu_1 \cdots \mu_i}$,
coming from the first term of Eq. (\ref{eq:20}),
can be reduced using Eq. (\ref{eq:15}) with
$I \to I^{(n)}$ 
\footnote{
\label{footI}
Of course, the tensors in the numerator of $I^{(n)}$ 
stay 4-dimensional.   
}.
However, there is still one important modification we have to take care of.
The 4-dimensional denominators appearing in Eq. (\ref{eq:13})
differ from the $n$-dimensional ones by an amount ${\tilde q}^2$
\bqa
\label{eq:24}
D_k=  \bar D_k-{\tilde q}^2\,.
\eqa
To compensate for this, the only replacement 
needed in Eq. (\ref{eq:15}), before applying it
to the $n$-dimensional case, is 
\bqa
\label{eq:25}
m^2_0 I_{m;\rho \cdots \tau}   \to
m^2_0 I^{(n)}_{m;\rho \cdots \tau}
     -I^{(n;\,2)}_{m;\rho \cdots \tau}
\eqa
Summarizing, any $n$-dimensional one-loop
amplitude can be written  
\bqa
\label{eq:26}
A^{(n)} = \sum_{m,j,i} 
I^{(n)}_{m,j;\,\bar \mu_1 \cdots \bar \mu_i}
\,\,S_{m,j}^{\bar \mu_1 \cdots \bar \mu_i}\,.
\eqa
After splitting the $n$-dimensional momenta according to
Eq. (\ref{eq:20}), $I^{(n)}_{m,j;\,\mu_1 \cdots \mu_i}$
can be evaluated with the help of Eq. (\ref{eq:15}), 
together with the replacement given in Eq. (\ref{eq:25}).
The additional integrals are always of the type 
$I^{(n;\,2 \ell)}_{m;\,\mu_1 \cdots \mu_h}$. Since, after all, they are also
4-dimensional tensors, it would be in principle possible to compute them 
with the help of Eq. (\ref{eq:15}). In practice, the direct computation
given in Appendix \ref{appa} 
is  more convenient when one is interested in taking the limit  
$\epsilon \to 0$.

To conclude, we notice that the technique of splitting loop tensors 
in 4-dimensional plus $\epsilon$-dimensional parts can help also 
outside the algorithm we are presenting here. For example, the method 
to reduce pentagon tensor integrals to box tensor ones 
presented in Ref. \cite{Denner:2002ii} relies on 4-dimensional
objects. 
However, a strategy for a possible extension to $n$-dimensions, 
which requires an explicit subtraction of soft and collinear 
divergences \cite{Beenakker:2001rj,Dittmaier:2003bc}, is outlined 
by the authors. 
Instead, as described above, via the splitting in Eq. (\ref{eq:20}), 
the $\epsilon$-dimensional part of the tensors can be computed 
separately; and the algorithm
of Ref. \cite{Denner:2002ii} directly applied to the remaining 
4-dimensional part of the tensor integrals 
($I^{(n)}_{4;\,\mu_1 \cdots \mu_i}$ in our notation) 
without any need of introducing a regulator $\Lambda$ in the 
intermediate stages of the calculation.
The only additional modification is the replacement 
in Eq. (\ref{eq:24}) for all the reconstructed denominators 
appearing in Ref. \cite{Denner:2002ii}. 

\section{Three-point tensors}
\label{sec:cfunctions} 
Eq. (\ref{eq:15}) cannot be applied when $m= 2$ 
because of the lack of a third 4-momentum $p_3$ 
to reconstruct denominators.
In this Section we derive a specific 
recursion relation for this case, which is valid for 
rank 2 and rank 3 three-point tensor integrals:
\footnote{\label{foot23}
The third index is in parentheses to remind that the equation 
is valid for rank 2 and 3.}
\bqa
\label{eq:3points}
I^{(n)}_{2;\,\mu \nu (\rho)}  &=& \frac{\beta}{2 \gamma}\, 
T^\prime_{\mu  \nu \lambda \sigma}\,\left\{J^{(n)\,\lambda\sigma}_{2;\,(\rho)}
\right\}
~-~\frac{1}{4 \,\gamma} t_{\mu \nu} 
\left\{m_0^2\,  I^{(n)}_{2;\,(\rho)} 
+I^{(n)}_{1;\,(\rho)}(0)-I^{(n;\,2)}_{2;\,(\rho)} \right\}\,, 
\eqa
where $J^{(n)\,\lambda\sigma}_{2;\,(\rho)}$ is the combination 
of one-loop integrals given in Eq. (\ref{eq:j}) and
\bqa
\label{eq:t2}
T^\prime_{\mu \nu \lambda \sigma} &=&
 g_{\mu \lambda} 
 \left(g_{\nu \sigma}-\frac{t_{\nu \sigma}}{2 \gamma} \right )
+g_{\nu \lambda} 
 \left(g_{\mu \sigma}-\frac{t_{\mu \sigma}}{2 \gamma} \right )
+\frac{g_{\lambda \sigma}}{2 \gamma}\, t_{\mu \nu}\,.
\eqa
Things get more complicated for higher rank tensors and
the general solution is given in Appendix \ref{appc}.

To derive the former recursion relation, we shall make use of 
the following 

\noindent 
{\underline {Theorem}:}
\bqa
\label{eq:theorem1}
&&\int d^n \bar q \frac{1}{\bar D_0\bar D_1\bar D_2} 
(q \cdot \ell_{3})^i= 0\,, \nl
&&\int d^n \bar q \frac{1}{\bar D_0\bar D_1\bar D_2} (q \cdot \ell_4)^i= 0\,,
~~~~~~\forall\, i=\,1,2,3\cdots \,.
\eqa
%

\noindent 
{\underline {Proof}:}
\bqa
\int d^n \bar q \frac{1}{\bar D_0\bar D_1\bar D_2} (q \cdot \ell_3)^i= 
\ell_3^{\mu_1} \cdots \ell_3^{\mu_i}\,I^{(n)}_{2;\mu_1\cdots \mu_i}\,.
\eqa
On the other hand, the tensor integral $I^{(n)}_{2;\mu_1\cdots \mu_i}$ 
admits a decomposition in terms of momenta $p_{1,2}$ and metric tensors
\bqa
\cdots p_{1\mu_k} \cdots p_{2\mu_j}\cdots g_{\mu_\ell \mu_h}\cdots\,.
\eqa
Then, as $(\ell_{3} \cdot p_{1,2})= 0$, all tensor structures containing 
$p_{1\mu_k}$ or $p_{2\mu_j}$ will vanish 
when contracted with $\ell_3^{\mu_k}$ or $\ell_3^{\mu_j}$. Analogously,
$g_{\mu_\ell \mu_h}$ cancels when 
contracted with $\ell_3^{\mu_\ell}\ell_3^{\mu_h}$ because $\ell_3^2= 0$. 
What proves the theorem. In the same way it can be shown the 
identity for $\ell_4$.

\noindent 
{\underline {Corollary}:}
\bqa
\label{eq:cor1}
\int d^n \bar q \frac{1}{\bar D_0\bar D_1\bar D_2} 
 (q \cdot \ell_{3})^2\,q_\rho 
~=~ \int d^n \bar q \frac{1}{\bar D_0\bar D_1\bar D_2} 
 (q \cdot \ell_4)^2\,q_\rho 
~=~ 0\,.
\eqa
It can be proved again performing an explicit tensor 
decomposition. 
This theorem allows to replace in the integrand of a rank 2 or 3
three-point integral
\bqa
\label{eq:to}
Q_{\mu}Q_{\nu}\,(q_\rho) ~\to~ \beta \left[D^\lambda q^\alpha \right]
\left\{ 
g_{\lambda \alpha} t_{\mu \nu} \right\}\,(q_\rho)  
- \gamma\, q^2 \left\{t_{\mu \nu} \right\}\,(q_\rho)\,. 
\eqa
Indeed, $Q_{\mu}Q_{\nu}$ in Eq. (\ref{eq:10}) read 
\bqa
Q_{\mu} Q_{\nu}= 
 (q \cdot \ell_3)^2 \ell_{4\mu}\ell_{4\nu}
+(q \cdot \ell_4)^2 \ell_{3\mu}\ell_{3\nu}
+(q \cdot \ell_3)(q \cdot \ell_4) t_{\mu\nu}\,.
\eqa
Then, the previous theorem guarantees that the first two terms give zero 
after integration, and the corollary that the same remains true 
when they are multiplied by {\em one and only one} additional 
integration momentum $q_\rho$.
Finally, Eq. (\ref{eq:to}) and the same steps as 
in Sections \ref{sec:1} and \ref{sec:2} 
result in the recursion relation (\ref{eq:3points}).

To conclude, we observe that similar methods can  
be also used to compute two-point tensors, as explicitly shown, 
for a particular case, at the beginning of next Section.
However, the case $m =1$ is so simple that we do not find 
any advantage with respect to standard reduction techniques.

\section{Rank one tensors}
\label{sec:rankone} 
As already observed at the end of Section \ref{sec:1}, 
Eq. (\ref{eq:15}) cannot be applied, as it stands, to reduce tensors 
of rank one.
In this Section we show how to cope with this situation.

\subsection{The $m=1$ case}
The standard Passarino-Veltman decomposition gives
\bqa
\label{onee}
I^{(n)}_{1;\,\mu}= \frac{p_1^\mu}{2\,p_1^2}
\left\{
 f_{10}\,I^{(n)}_{1}
        +I^{(n)}_{0}(1)
        -I^{(n)}_{0}(0)
\right\}\,.
\eqa
The same result can be derived extending our method.
We can write
\bqa
\label{eq:spb}
p_1= \ell_1+\frac{\ell_2}{2}~~
{\rm with}~~ \ell_{1,2}^2= 0~~{\rm and}~~(\ell_1 \cdot \ell_2) 
= 2\,(\ell_1 \cdot p_1) = (\ell_2 \cdot p_1) = p_1^2\,,
\eqa
what corresponds to the second case of 
Footnote \ref{foot} with $\alpha_1= 1/2$.
The massless 4-vectors $\ell_{1,2}$ in Eq. (\ref{eq:spb}) 
always exist. For example,
if $p_1^2  >  0$,  $\ell_1= (M/2,-M/2,0,0)$ and $\ell_2= (M,M,0,0)$  
in the frame where $p_1= (M,\vec 0)$. 
The corresponding
4-vectors $ \ell_{3,4}$ are defined as in Eq. (\ref{eq:l34}).
As besides $I^{(n)}_{1;\,\mu}$ must be proportional to $p_{1\mu}$, 
Eqs. (\ref{eq:5a}) and (\ref{eq:spb}) imply 
\bqa
\label{eq:theorem2}
\int d^n \bar q \frac{1}{\bar D_0\bar D_1} (q \cdot \ell_{3,4})
& ~=~ & 0\,, \nl
\int d^n \bar q \frac{1}{\bar D_0\bar D_1}\,2\,(q \cdot \ell_1)
& ~=~ &\int d^n \bar q \frac{1}{\bar D_0\bar D_1}(q \cdot \ell_2) \nl 
& ~=~ &\int d^n \bar q \frac{1}{\bar D_0\bar D_1} (q \cdot p_1)\,. 
\eqa
Using them and Eq. (\ref{eq:8}) one obtains
\bqa
\int d^n \bar q \,\frac{q^\mu}{\bar D_0\bar D_1}~=~
\frac{1}{\gamma}\int d^n \bar q \,
\frac{2\,(q\cdot p_1)\,p^\mu_1}{\bar D_0\bar D_1}\,,
\eqa
and Eq. (\ref{onee}).

\subsection{The $m=2$ case}
Using Eq. (\ref{eq:8}) and the Theorem in Eq. (\ref{eq:theorem1}) one 
can show that 
\bqa
\int d^n \bar q \,\frac{q^\mu}{\bar D_0\bar D_1\bar D_2}~=~
\frac{\beta}{\gamma}
\int d^n \bar q \,\frac{D^\mu}{\bar D_0\bar D_1\bar D_2}\,;
\eqa
so that
\bqa
I^{(n)}_{2;\,\mu}= \frac{\beta}{\gamma} J^{(n)}_{2;\,\mu}\,,
\eqa
where $J$ is defined in Eq. (\ref{eq:j}).

\subsection{The $m=3$ case}
We use again Eq. (\ref{eq:8}) to write
\bqa
\label{eq:4point1}
I^{(n)}_{3;\,\mu}=
\int \frac{d^n \bar q}{\bar D_0\bar D_1\bar D_2\bar D_3}
\left[
 \frac{\beta\,D_\mu}{\gamma}
-\frac{\ell_{3\mu}(q \cdot \ell_4)+\ell_{4\mu}(q \cdot \ell_3)}{2\,\gamma}
\right]\,.
\eqa
Multiplying and dividing $(q \cdot \ell_{3,4})$ by 
$(p_3 \cdot \ell_{4,3})$ we can express
\bqa
(q \cdot \ell_{3,4})= 
\frac{{\rm Tr}[\rlap/ \ell_1\rlap/ q \rlap/ \ell_2 
               \rlap/ p_3\,\omega^-]}{(p_3 \cdot \ell_{4,3})}\,,
\eqa 
where the trace containing $\gamma_5$ 
vanishes upon integration because it is proportional to 
the totally antisymmetric tensor 
$\epsilon(\ell_1,q,\ell_2,p_3)$. Then, the following substitution 
is allowed in the integrand of Eq. (\ref{eq:4point1}) 
(where we also use Eq. (\ref{eq:8}))
\bqa
(q \cdot \ell_{3,4})~\to~\frac{1}{(p_3 \cdot \ell_{4,3})}
\left[
\beta\, p_{3}^{\lambda}D_\lambda -\gamma (q \cdot p_3)
\right]\,,
\eqa
from which the desired result follows 
(using Eqs. (\ref{eq:j}) and (\ref{eq:9a}))
\bqa
I^{(n)}_{3;\,\mu}  &=& \frac{\beta}{\gamma}\,J^{(n)}_{3;\,\mu} 
~+~\frac{1}{4}\,\left[
\frac{\ell_{3\mu}}{(p_3 \cdot \ell_3)}+
\frac{\ell_{4\mu}}{(p_3 \cdot \ell_4)}
\right]\nl
&\times& \left\{f_{30} I^{(n)}_{3}      
             +I^{(n)}_{2}(3)
             -I^{(n)}_{2}(0)
 -\frac{2 \beta}{\gamma} p_{3}^{\lambda}\,J^{(n)}_{3;\,\lambda}
\right\}\,.
\eqa

\subsection{The $m > 3$ case}
Contracting Eq. (\ref{eq:8}) with $p_{3,4}$ one can write $q_\mu$ 
as a function of $(q \cdot p_i)$, $i= 1, \cdots , 4$:
\bqa
\label{eq:mgt3}
q_\mu &=& \frac{\beta}{\gamma} D_\mu 
+ \frac{\ell_{3\mu}\ell_{4\alpha}-\ell_{3\alpha}\ell_{4\mu}}{2\,\delta}
\left\{ 
     p_3^\alpha 
       \left( 2\, (q\cdot p_4) - \frac{2\beta}{\gamma} p_{4\lambda} D^\lambda
       \right)
    -p_4^\alpha 
       \left( 2\, (q\cdot p_3) - \frac{2\beta}{\gamma} p_{3\lambda} D^\lambda
       \right)
\right\},\nl
\delta &=& (\ell_3\cdot p_4)(\ell_4\cdot p_3)
          -(\ell_3\cdot p_3)(\ell_4\cdot p_4)\,.
\eqa
Then, after reconstructing the denominators
and integrating, one gets
\bqa
I^{(n)}_{m;\,\mu}  &=& \frac{\beta}{\gamma}
J^{(n)}_{m;\,\mu}
+ \frac{\ell_{3\mu}\ell_{4\alpha}-\ell_{3\alpha}\ell_{4\mu}}{2\,\delta}
\left\{
 p_3^\alpha
\left[f_{40}  I^{(n)}_{m}      
             +I^{(n)}_{m-1}(4)
             -I^{(n)}_{m-1}(0)
 -\frac{2 \beta}{\gamma} p_{4\lambda}\,J^{(n)\,\lambda}_{m}
 \right] \right. \nl
&-& p_4^\alpha \left.
\left[f_{30}  I^{(n)}_{m}      
             +I^{(n)}_{m-1}(3)
             -I^{(n)}_{m-1}(0)
 -\frac{2 \beta}{\gamma} p_{3\lambda}\,J^{(n)\,\lambda}_{m}
 \right]
\right\}\,.
\eqa
Note that $|\delta|^2$ is proportional to the Gram determinant
of the 4-momenta $\ell_1,\ell_2, p_3, p_4$. The appearance of
inverse square roots of Gram determinants is a peculiarity of
our formalism and it will be discussed at length in Section \ref{sec:coll}.

We close this Section by observing that Eq. (\ref{eq:mgt3}) could be used 
instead of Eq. (\ref{eq:13}) when $m>3$ to derive 
a recursion relation alternative to Eq. (\ref{eq:15}). 
In fact nothing prevents from multiplying  by an arbitrary 
number of 4-vectors $q_\rho \cdots q_\tau$ before integrating.
The reason why we prefer Eq. (\ref{eq:13}) is because it involves
only three out of the $m$ external 4-momenta, while
a fourth momentum is necessary to write down
Eq. (\ref{eq:mgt3}). This has important consequences 
when studying collinear or coplanar configurations, 
as we shall see in next Section. 
 
\section{Study of exceptional configurations}
\label{sec:coll} 
We are interested in the behaviour of our formulae 
at the edges of the phase-space, where
two or more momenta can become nearly linearly dependent. 
We shall first show that
only square roots of the two Gram determinants 
\bqa
\Delta & = & -\Delta_{12} ~~~=~~~  (p_1 \cdot  p_2)^2-p_1^2 p_2^2 \nl
\Delta_{123}&=& 
           2\,(\ell_1 \cdot  \ell_2)(\ell_1 \cdot p_3)(\ell_2 \cdot p_3)
           -p_3^2\,(\ell_1 \cdot  \ell_2)^2
\eqa
appear in the denominators of Eq. (\ref{eq:15}).
Furthermore, the only occurrence of a square root of a
rank four Gram determinant is in Eq. (\ref{eq:mgt3}).
These two facts make our approach
numerically more stable than conventional methods, in which
the scalar coefficients of the tensor decomposition develop
poles proportional to $1/\Delta_{12}$, $1/\Delta_{123}$ or 
$1/|\delta|^2$ at each step of the reduction.
Secondly, we shall argue that Eq. (\ref{eq:15}) is
such that the numerical cancellations 
occurring among the tensor loop functions in the limit of exceptional 
momenta are kept as local as possible. Otherwise stated,
each of the three terms of Eq. (\ref{eq:15})
is separately well behaved when $\Delta_{12} \to 0$ or $\Delta_{123} \to 0$.
Although this cannot solve by itself all problems of 
numerical inaccuracy, it helps in decreasing 
the values of $\Delta_{12}$ and $\Delta_{123}$
for which approximations to Eq. (\ref{eq:15}) should be used.
Thirdly, we shall show how to deal with 
configurations with {\em exactly} zero
Gram determinants and give general prescriptions on how to
cure the numerical instabilities occurring near the zeros
of $\Delta_{12}$ and $\Delta_{123}$.

We should then investigate all possible denominators appearing 
in Eq. (\ref{eq:15}) in the limit of exceptional momenta.
We start considering the 4-vectors $\ell_{1,2}$ in Eq. (\ref{eq:4}).
Inserting $\alpha_{1,2}$ in the definition of $\beta$, one finds 
\bqa
\beta= \pm \frac{p_1^2}{2 \alpha_1 \sqrt{\Delta}}\,,
\eqa
where the sign depends on the sign in $\alpha_1$ 
(see Footnote \ref{foot}).
Despite of this, $\ell_{1,2}$ remain well behaved 
in the limit $\Delta_{12} \to 0$.
In fact, splitting the external momentum $p_2$ as follows
\bqa
\label{eq:split2}
p_2= \eta p_1+ (p_2-\eta p_1) \equiv \eta p_1 + \phi {\hat n}\,,
\eqa
and choosing $\eta$ and $\phi$ such that $(p_1 \cdot \hat n) = 0$ and
$\hat n^2=-p_1^2$, one gets
\bqa
p_2= \eta p_1 + \frac{\sqrt{\Delta}}{p_1^2}\,\hat n\,,
\eqa
with
\bqa
\eta  =  \frac{(p_1 \cdot p_2)}{p_1^2}\,~~~{\rm and}~~~
   \hat n^\mu = \frac{1}{\sqrt{\Delta}}
\left[ p_1^2\, p_2^\mu - (p_1  \cdot p_2)\, p_1^\mu
\right]\,.
\eqa
Therefore, in terms of $p_1$ and $\hat n$, we get well defined expressions
even when $\Delta_{12}\to 0$:
\bqa
\label{eq:elles}
\ell_1= \frac{1}{2} (p_1 \mp \hat n)\,,~~~~
\ell_2= \frac{1}{2 \alpha_1} (p_1 \pm \hat n)\,.
\eqa
If needed, one can use higher numerical accuracy just in the computation 
of $\hat n^\mu$, or choose a particular frame to stabilise the result. 
For example, for time-like $p_1$, one takes
$\hat n^\mu = M\, (0, \vec{p}_2/|\vec{p}_2|)$ when $p_1^\mu = (M,\vec 0)$.
Finally, when both $p_i^2$ vanish,  
$\alpha_{1,2}= 0$ as noticed in Footnote \ref{foot},
and $\ell_{1,2} \equiv p_{1,2}$.
When only one $p_i^2$ is zero, $\ell_{1,2}$ are still well defined
provided $(p_1 \cdot p_2)\ne 0$, but this can also be cured, 
as discussed later.

Let us now analyse the terms $\beta$, $1/\gamma$, 
$1/(\ell_3 \cdot p_3)$ and $1/(\ell_4 \cdot p_3)$ in Eq. (\ref{eq:15}) 
in turn, where
the last two quantities are hidden
in the definition of the 
rank three tensor $T_{\mu\nu\lambda}$.
We start considering the case $p_i^2 \ne 0$. Then $\beta$ 
is proportional to $1/\sqrt{\Delta}$, but the combination 
of tensor loop functions multiplying 
$\beta$ in Eq. (\ref{eq:15}) is always such that the product
is well behaved in the limit
$\Delta_{12}\to 0$.
The proof is simple and follows from the fact that these terms
come from $D^\mu$ defined in Eq. (\ref{eq:8}).
Indeed, it can be shown using Eq. (\ref{eq:elles}) that
\bqa
\label{eq:beh12}
D^\mu = \mp 2  \frac{\sqrt{\Delta}}{p_1^2} 
\left[(q \cdot \hat n)\hat n^\mu -(q \cdot p_1) p_1^\mu \right]\,,
\eqa
thus compensating the $1/\sqrt{\Delta}$ pole coming from $\beta$.
Of course, we cannot simplify this pole analytically, because by doing so 
we cannot express $D^\mu$ back in terms of denominators. 
However, Eq. (\ref{eq:beh12}) shows that well behaved combinations 
of loop functions naturally arise in our method. 
This is in contrast with the well behaved groupings of Ref.
\cite{Campbell:1996zw}, result of a very complicated compensation 
of logarithms and di-logarithms in groupings obtained by differentiating
with respect to external parameters or by developing scalar integrals in 
$6+\epsilon$ or higher dimensions. 
Obviously, such a compensation has also to occur {\em after} 
integrating over $d^4q$. The nice feature of our approach is that 
this works {\em before} performing the actual integration, 
what keeps things much simpler.
In addition, in our case the cancellations in the numerator should compensate
a factor ${1}/{\sqrt{\Delta}}$, and not ${1}/{{\Delta}}$.
Finally, when at least one $p_i$ is massless, $\beta$ is simply 1.

Next we concentrate on the zeros of $\gamma$. Since
\bqa
\gamma= \frac{p_1^2 p_2^2 }{(p_1 \cdot p_2) \mp \sqrt{\Delta}}\,,
\eqa
there is no problem for massive $p_{1,2}$.
On the other hand, when $p_{1,2}$ are both massless, 
$\gamma= 2 {(p_1 \cdot p_2)}$ can vanish. 
However, such configurations correspond to 
true collinear singularities of the amplitude, which
are cut away in physical observables. 
When $p_1^2 = 0$ and $p_2^2 \ne 0$ 
($p_1^2 \ne 0$ and $p_2^2= 0$), $\gamma$ 
can become zero at the edges of the phase-space 
because $\gamma= 2 {(p_1 \cdot p_2)} = 0$
is a zero of the Gram determinant 
$\Delta_{p_{1}-p_{2},p_2}$ ($\Delta_{p_{1},p_{2}-p_1}$). 
For loop functions with $m \ge 3$, one simply
renames $p_{1,3}$ assigning the massless 4-momentum to $p_3$.
On the other hand, if $m = 2$ and to fix things $p_1^2 = 0$,
one redefines $\ell_{1,2}$ in Eq. (\ref{eq:3}) using 
$p_1-p_2$ and $p_2$ instead of $p_1$ and $p_2$
in order to move this pole to $\beta$. 
The latter is important because as we have seen, $\beta$ behaves 
as the inverse square root of a Gram determinant, 
while poles in ${1}/{\gamma^2}$ are present everywhere 
in our formulae.
The net effect of this choice for $\ell_{1,2}$ 
is the replacement
\bqa
  I^{\sigma}_{m-1;\rho \cdots \tau}(1) \to
  I^{\sigma}_{m-1;\rho \cdots \tau}(1)
 -I^{\sigma}_{m-1;\rho \cdots \tau}(2)
\eqa
in Eq. (\ref{eq:15}).

We turn to $1/(\ell_3 \cdot p_3)$ 
and $1/(\ell_4 \cdot p_3)$.
They are both proportional to $1/\sqrt{\Delta_{123}}$.
In fact, 
\bqa
|(\ell_4 \cdot p_3)|^2 &=& |(\ell_3 \cdot p_3)|^2 
~=~
\bar v(\ell_1)\, \rlap/p_3\,       \omega^-\, u(\ell_2)\,
\bar v(\ell_2)\, \rlap/p_3\,       \omega^-\, u(\ell_1)
~=~
 \bar v(\ell_1)\, \rlap/p_3\,\rlap/\ell_2 \rlap/p_3\,
                                 \omega^-\, u(\ell_1) \nl
&=& \frac{1}{2}{\rm Tr} [\rlap/ \ell_1\rlap/ p_3 \rlap/ \ell_2 \rlap/ p_3]
 ~=~ \frac{2}{(\ell_1 \cdot \ell_2)}\Delta_{123}\,.
\eqa
This completes the proof of that only square roots of Gram 
determinants appear at each step of our reduction procedure. 

Now, we shall show that also the last two terms of Eq. (\ref{eq:15}) 
contain only well behaved combinations of loop functions in the limits
$\Delta_{12} \to 0$ and $\Delta_{123} \to 0$.
To prove this, we split the 4-vector $p_3$ in a similar way as we did 
with $p_2$ in Eq. (\ref{eq:split2}) 
\bqa
\label{eq:split3}
p_3 = \eta_1 \ell_1 + \eta_2 \ell_2 + \phi \hat m\,.
\eqa
Choosing $\eta_1$, $\eta_2$ and $\phi$ such that
$(\ell_1 \cdot \hat m) = (\ell_2 \cdot \hat m) = 0$ and
$\hat m^2 = - (\ell_1 \cdot \ell_2 )^2$, one gets
\bqa
\eta_1 &=& \frac{(\ell_2 \cdot p_3)}{(\ell_1 \cdot \ell_2)}\,,~~~
\eta_2 ~=~ \frac{(\ell_1 \cdot p_3)}{(\ell_1 \cdot \ell_2)}\,,~~~
\phi    ~=~ \frac{\sqrt{\Delta_{123}}}{(\ell_1 \cdot \ell_2)^2}\,, \nl
\hat m^\mu &=&  \frac{(\ell_1 \cdot \ell_2)}{\sqrt{\Delta_{123}}}
\left[
 (\ell_1 \cdot \ell_2)\,{p_3}^\mu 
-(\ell_2 \cdot    p_3)\,{\ell_1}^\mu 
-(\ell_1 \cdot    p_3)\,{\ell_2}^\mu 
\right]\,.
\eqa
Then
\bqa
(\ell_{3,4} \cdot p_3) = 
\frac{\sqrt{\Delta_{123}}}{(\ell_1 \cdot \ell_2)^2}
(\ell_{3,4} \cdot \hat m)\,,~~~
\eqa
implying that $T_{\mu \nu}$, defined in Eq. (\ref{eq:12}) and 
depending only on the ratio between
$(\ell_{3} \cdot p_3)$ and $(\ell_{4} \cdot p_3)$,
behaves smoothly in the limit $\Delta_{123} \to 0$.
This tells us that the only singularity in the 
second term of Eq. (\ref{eq:15}) can come from $1/\gamma$, but this can 
be cured as explained above.
To conclude, the last term of Eq. (\ref{eq:15}) is proportional to 
$T_{\mu\nu\lambda} \sim \frac{1}{\sqrt{\Delta_{123}}}$.
Expressing $\ell_{1,2}$ in Eq. (\ref{eq:split3})
in terms of $p_{1,2}$ one obtains
\bqa
\label{eq:split3a}
p_3 = 
  \left[\frac{2\beta}{\gamma} (r_2 \cdot p_3)\right] p_1 
 +\left[\frac{2\beta}{\gamma} (r_1 \cdot p_3)\right] p_2 
+ \frac{\sqrt{\Delta_{123}}}{(\ell_1 \cdot \ell_2)^2}\, \hat m\,.
\eqa
Then, using Eq. (\ref{eq:split3a}) the coefficient of
$q^\lambda \,T_{\mu\nu\lambda}$ in Eq. (\ref{eq:13}) can be written as
\bqa
-\frac{1}{2 \gamma}(q \cdot \hat m)\,
 \frac{\sqrt{\Delta_{123}}}{(\ell_1 \cdot \ell_2)^2}\,.
\eqa
Therefore, the last combination of loop functions in Eq. (\ref{eq:15})
must combine in such a way that the pole 
${1}/{\sqrt{\Delta_{123}}}$ coming from the tensor gets compensated.

When $\Delta_{12}$ or $\Delta_{123}$ are {\em exactly} zero 
one has to rely on a different strategy.
Let us concentrate on the case when, for example, 
$p_2= \lambda p_1$ {\em exactly}.
If $m$ is large enough, one simply chooses within the set 
$\{p_j\}$ a different
subset of three independent 4-momenta to perform the reduction. 
$\bar D_2$ acts then as a {\em spectator} denominator.
When due to cancellations of denominators one is left with tensors 
integrals
\bqa
\int d^n \bar q \frac{q_\mu \cdots q_\nu}{\bar D_0 \bar 
D_1 \bar D_2 \bar D_3}\,,
\eqa
one switches to the reduction valid for
3-point functions. In fact, when $p_2= \lambda p_1$, 
\bqa
 \int d^n \bar q \frac{q_\mu \cdots q_\nu}{\bar D_0 \bar D_1 \bar D_2 \bar D_3}
~~~{\rm and}~~~
 \int d^n \bar q \frac{q_\mu \cdots q_\nu}{\bar D_0 \bar D_1 \bar D_3}\,
\eqa
share the same tensor basis built up in terms of $p_{1,3}$ 
and metric tensors.
At the end, only 2-point like tensors remain 
to be further reduced with the standard techniques of 
Ref. \cite{Passarino:1978jh}.
The same procedure also works 
when three or more momenta become linearly dependent.
With the outlined method, reliable approximations 
can be easily obtained near the zeros of the Gram determinants, where
numerical cancellations occur among the loop functions of
each well behaved combination. 
When, for example, $\Delta_{12} \ll 1$, 
one simply starts the reduction using
$\bar D_2 = (\bar q + \lambda p_1)^2-{m_2}^2$.
This is completely equivalent to the Taylor expansion 
performed in Ref. \cite{Campbell:1996zw} to extract 
the constant terms in each grouping. 
The advantage of our approach is that we do not need to explicitly
develop tensor functions. Just at the end, when everything is 
reduced to scalar $(4+\epsilon)$-dimensional loop integrals, 
only these need to be computed 
precisely, also in the regime of vanishing Gram determinants. 
This can be done, for example, as explained in Refs. 
\cite{Tarasov:1996br,Duplancic:2003tv,Binoth:2002xh,Campbell:1996zw}.

A problem which remains to be solved is the determination 
of the maximal values of $\Delta_{12}$,
$\Delta_{123}$ and $\delta$ in Eq. (\ref{eq:mgt3}) below which
the approximations to Eq. (\ref{eq:15}) should be used.
As in Ref. \cite{Campbell:1996zw} such values can be only found 
performing dedicated numerical studies, and it is then difficult 
to give general prescriptions.
We do not want to get deeply involved into the subject here, but 
just mention that by taking advantage of the fact 
that our reduction procedure takes place {\em before} integration,
tests on the numerical stability 
of the formalism are possible without even evaluating the loop integrals.
For example, for any given arbitrary 4-vector $q$ 
the {\em integrands} on the r.h.s. of Eq. ({\ref{eq:15}) should add up in such 
a way that at the end of the recursive algorithm 
the result is numerically equivalent to
\bqa
\frac{q_\mu q_\nu q_\rho \cdots q_\tau}{D_0 D_1 \cdots D_m}\,.
\eqa
We have performed such a check on tensor integrals up to rank 4.

\section{Summary}
\label{sec:Summary}
We have presented a method to compute numerically and recursively
tensor integrals appearing in one-loop calculations, and relevant 
for the next generation of $pp$ and $e^+e^-$ colliders.
The treatment is applicable 
irrespective of the number of external legs 
to any configuration of internal and/or external 
variables, and only requires the knowledge 
of the standard set of scalar one-loop integrals.
We distinguish the cases of 3-point tensor integrals 
(Section \ref{sec:cfunctions}), 
as well as of rank 1 (Section \ref{sec:rankone}), 
which are treated separately with 
similar techniques to those used in the general case
(Sections \ref{sec:1} and \ref{sec:2}).
Singular kinematical configurations are analised in 
detail (Section \ref{sec:coll}), finding a smoother behaviour than 
in other approaches. 
In addition, we have studied all possible sources 
of numerical instabilities, giving general prescriptions 
on how to cure them.
A code implementing the proposed method will be 
made available in the near future.

\section*{Acknowledgments}
R. P. would like to thank all members of the Department of Theoretical 
Physics of the University of Granada for the warm hospitality.
This work is supported by the European Union 
Community's Human Potential Programme 
under contract HPRN-CT-2000-00149 
Physics at Colliders, and by
MECD under contact SAB2002-0207. 

\appendix
\section{The vectors $\ell_{3,4}$\label{app:1}}
The 4-vectors $\ell_{3,4}$ defined in Eq. (\ref{eq:l34}) 
enjoy useful properties. By using the Dirac equation
and the completeness relations for massless spinors
one immediately derives 
\bqa
(\ell_{3,4} \cdot \ell_{1,2}) &=& 0\,, \nl 
\ell_3^2&=& 
\bar v(\ell_1)\, \gamma^\mu\,       \omega^-\, u(\ell_2)\,
\bar v(\ell_1)\, \gamma_\mu\,       \omega^-\, u(\ell_2) \nl
&=&  \frac{1}{\bar v(\ell_2)\,\rlap/b\,u(\ell_1)}\,
   \bar v(\ell_1)\, \gamma^\mu\,       \omega^-\, u(\ell_2)\,
   \bar v(\ell_2)\,\rlap/b                     \, u(\ell_1)\,
   \bar v(\ell_1)\, \gamma_\mu\,       \omega^-\, u(\ell_2)\nl
&=&  \frac{1}{\bar v(\ell_2)\,\rlap/b\,u(\ell_1)}\,
   \bar v(\ell_1)\, \gamma^\mu\,\rlap/\ell_2
         \,\rlap/b\, \rlap/ \ell_1 \,\gamma_\mu\,\omega^-\, u(\ell_2)\nl
&=&  \frac{-2}{\bar v(\ell_2)\,\rlap/b\,u(\ell_1)}\,
    \bar v(\ell_1)\, \rlap/\ell_1
         \,\rlap/b\, \rlap/ \ell_2\,\omega^-\, u(\ell_2)~=~0\,, \nl
\ell_4^2 &=& 
\bar v(\ell_2)\, \gamma^\mu\,       \omega^-\, u(\ell_1)
\bar v(\ell_2)\, \gamma_\mu\,       \omega^-\, u(\ell_1)
~=~0\,,
\eqa
where $b$ is an arbitrary 4-vector different from $\ell_{1,2}$. 
Furthermore, 
\bqa
(\ell_3\cdot \ell_4)&=&
\bar v(\ell_1)\, \gamma_\mu\,       \omega^-\, u(\ell_2)\,
\bar v(\ell_2)\, \gamma^\mu\,       \omega^-\, u(\ell_1)
~=~
 \bar v(\ell_1)\, \gamma_\mu\,\rlap/\ell_2 \gamma^\mu\,
                                 \omega^-\, u(\ell_1) \nl
&=& \frac{1}{2}{\rm Tr} [\rlap/ \ell_1 \gamma_\mu  \rlap/ \ell_2 \gamma^\mu]
~=~-4\,(\ell_1 \cdot \ell_2)\,, \nl
(q \cdot \ell_3)(q \cdot \ell_4) &=&
\bar v(\ell_1)\, \rlap/q\,       \omega^-\, u(\ell_2)\,
\bar v(\ell_2)\, \rlap/q\,       \omega^-\, u(\ell_1)
~=~
 \bar v(\ell_1)\, \rlap/q\,\rlap/\ell_2 \rlap/q\,
                                 \omega^-\, u(\ell_1) \nl
&=& \frac{1}{2}{\rm Tr} [\rlap/ \ell_1\rlap/ q \rlap/ \ell_2 \rlap/ q]
~=~4\,(q \cdot \ell_1)(q \cdot \ell_2) -2\,q^2\,(\ell_1 \cdot \ell_2)\,, \nl
(q \cdot \ell_3)(q \cdot \ell_3) &=& \frac{1}{(b \cdot \ell_4)}
   \bar v(\ell_1)\, \rlap/q\,       \rlap/\ell_2\,
                 \, \rlap/b\,       \rlap/\ell_1\,
                 \, \rlap/q\,       \omega^-\, u(\ell_2)\,  \nl
&=& \frac{2}{(b \cdot \ell_{4})}
\left\{ [q^2 (\ell_1 \cdot \ell_2)-2
           (q \cdot \ell_1)(q \cdot \ell_2)]
           (b \cdot \ell_{3}) \right. \nl
    &+&\!\!\!\!\!\!\left.2\, 
         [(q \cdot \ell_1)(b \cdot \ell_2)
         -(q \cdot b)(\ell_1 \cdot \ell_2)
         +(q \cdot \ell_2)(\ell_1 \cdot b)]
          (q \cdot \ell_{3}) 
            \right\}\,.
\eqa
In the same way one computes $(q \cdot \ell_4)(q \cdot \ell_4)$.

For numerical applications one needs
$\ell_{3,4}$ in terms of  $\ell_{1,2}$.
Given $\ell_i^\mu=(\ell_{i0},\ell_{ix},\ell_{iy},\ell_{iz})$ 
and using \cite{Kuijf:1991kn}:
\bqa
\sigma^0&=&
\left(
\begin{tabular}{cc}
 $1$ & $0$ \\
 $0$ & $1$ \\
\end{tabular}
\right)
\,,~~~
\sigma^1=
\left(
\begin{tabular}{cc}
$ 0$ & $-1$\\
$-1$ & $0$ \\
\end{tabular}
\right)
\,,~~~
\sigma^2=
\left(
\begin{tabular}{cc}
$ 0$ & $i$ \\
$-i$ & $0$ \\
\end{tabular}
\right)
\,,~~~
\sigma^3=
\left(
\begin{tabular}{cc}
 $-1$& $0$ \\
 $0$ & $1$ \\
\end{tabular}
\right)\,, \nl
\gamma^0 &=&
\left(
\begin{tabular}{cc}
 0 & $\sigma^0$ \\
 $\sigma^0$ & 0 \\
\end{tabular}
\right)
\,,~~~
\gamma^k =
\left(
\begin{tabular}{cc}
 0 & $-\sigma^k$ \\
 $\sigma^k$ & 0 \\
\end{tabular}
\right)
\,,~~~
\gamma^5 =
\left(
\begin{tabular}{cc}
 $\sigma^0$ & 0 \\
 0         & $-\sigma^0$ \\
\end{tabular}
\right)
\,, \\
\bar v(\ell_i)\, \omega^+ &=& ({b_i},{c^-_i}, 0 , 0)\,,~~~
\omega^- u(\ell_i)  =
\left(
\begin{tabular}{l}
$0$  \\ $0$ \\ ${b_i}$ \\ $ {c^+_i}$ 
\end{tabular}
\right)\,,~~~
{b_i}=\sqrt{\ell_{i0}+\ell_{iz}}\,,~~~ 
{c^{\pm}_i}=\frac{\ell_{ix}\pm i \ell_{iy}}{b_i}
\nonumber\,,
\eqa
the desired result follows
\bqa
\begin{tabular}{ll}
$\ell_{30} =     {b_1}{b_2}  +  {c^-_1}{c^+_2}$\,, &  
$\ell_{40} =     {b_2}{b_1}  +  {c^-_2}{c^+_1}$\,,   \\ 
$\ell_{3x} =     {b_1}{c^+_2}  +  {c^-_1}{b_2}$\,, &  
$\ell_{4x} =     {b_2}{c^+_1}  +  {c^-_2}{b_1}$\,,   \\ 
$\ell_{3y} = i(  {c^-_1}{b_2}  -  {b_1}{c^+_2}$)\,, &  
$\ell_{4y} = i(  {c^-_2}{b_1}  -  {b_2}{c^+_1}$)\,,   \\ 
$\ell_{3z} =     {b_1}{b_2}  -  {c^-_1}{c^+_2}$\,, &
$\ell_{4z} =     {b_2}{b_1}  -  {c^-_2}{c^+_1}\,.$   \\ 
\end{tabular}
\eqa

\section{Extra integrals 
\label{appa}}
In this Appendix we compute the extra integrals
appearing in the $n$-dimensional version of the proposed 
reduction method
\bqa
\label{eq:app1}
I^{(n;\,2 \ell)}_{m;\,\mu_1 \cdots \mu_{2s}} = 
\int d^n\bar q ~{\tilde q}^{2\ell} ~\frac{ q_{\mu_1} 
\cdots q_{\mu_{2s}}}{\bar D_0 \cdots\bar  D_m}\,,
\eqa
where $\ell > 0$ and $2s$ are non-negative integers. 
It is convenient to define a new index 
$d=\ell+s+1-m$ to classify
the integrals according to their dimensionality in powers of 
$[mass]^2$.
One can convince him/herself that tensor integrals have 
a non-zero contribution at $\ord(1)$, which is the order 
we are interested in, only when
$d \ge 0$. 
For example,
\bqa
I^{(n;\,2)}_{2;\,\mu}= \frac{i \pi^2}{6} (p_{1\mu}+p_{2\mu}) 
               + \ord(\epsilon)\,.
\eqa
Their general expression can be easily written for $d \le 0$:
\bqa
I^{(n;\,2(m-1+d-s))}_{m;\,\mu_1 \cdots \mu_{2s}} =
\left\{
\begin{tabular}{ll}
$-i \pi^2\,
\frac{g_{\mu_1 \cdots \mu_{2s}}}{2^s}\,\,
\frac{\Gamma(m-s-1)}{\Gamma(m+1)} + \ord(\epsilon)$ & when $d = 0$ , \\
$0$                                               & when $d < 0$ , 
\end{tabular}
\right.
\eqa
where the symmetric combination $g_{\mu_1 \cdots \mu_{2s}}$
is as in Eq. (\ref{eq:21}), but for 4-dimensional metric tensors.

Let us discuss the scalar case ($s=0$) in more detail.
Decomposing the integration
\cite{Mahlon:1993fe}
\bqa
\label{eq:app2}
\int d^n \bar q= \int d^4q\,d^\epsilon \mu~~~~~(\tilde{q}^2= -\mu^2)\,
\eqa
and after using Feynman parametrization, one gets
\bqa
\label{eq:app3}
I^{(n;\,2 (m-1+d))}_{m} &=& (-)^{2 m+d}\, i \pi^2
\frac{\pi^{\epsilon/2}}{\Gamma(\epsilon/2)} 
\Gamma\left(m-1+d+\frac{\epsilon}{2}\right)
\Gamma\left(-d-\frac{\epsilon}{2}\right) \nl
&\times& \int [d \alpha]_m~{\cal X}_m^{(d+\frac{\epsilon}{2})}\,,
\eqa
with
\bqa
\label{eq:app4}
\int [d \alpha]_m &=& \int_0^{\infty} d\alpha_0 \cdots d\alpha_m~ 
\delta(1-\sum_{k= 0}^m \alpha_k)\,,~~~{\cal X}_m = P_m^2+M_m^2 \nl
P_m &=& \sum_{k= 1}^m \alpha_k p_k\,,
~~~M_m^2 = \alpha_0 m_0^2+\sum_{k= 1}^m \alpha_k (m_k^2-p_k^2) \,.
\eqa

We explicitly compute here one-loop integrals coming from tensors with
rank at most equal to the number of denominators. This is enough for
most practical calculations and
gives rise to only one possibility with $d= 1$, namely $m= 1$.
A straightforward calculation for this integral, gives
\bqa
\label{eq:app5}
I^{(n;\,2)}_{1} &\equiv& 
\int d^n\bar q \,\frac{{\tilde q}^2}{\bar D_0 \bar D_1} = 
-i \frac{\pi^2}{2}\left[ m_1^2+m_0^2-\frac{p_1^2}{3}\right] 
+ \ord(\epsilon)\,.
\eqa
When $m_0^2=m_1^2=p_1^2=0$ the integral vanishes 
in dimensional regularization. 

Furthermore, all scalar integrals 
with $d = 0$ are easily evaluated by 
observing that  $\int [d \alpha]_m = {1}/{\Gamma(m+1)}$:
\bqa
\label{eq:app6}
I^{(n;\,2(m-1))}_{m} =
-i  \pi^2 \frac{\Gamma(m-1)}{\Gamma(m+1)} + \ord(\epsilon)
~~~~{\rm when}~~d = 0\,.
\eqa

When $d=-1$, Eq. (\ref{eq:app3}) gives
\bqa
\label{eq:app7}
I^{(n;\,2 (m-2))}_{m} &=& -i \pi^2
\frac{\pi^{\epsilon/2}}{\Gamma(\epsilon/2)} 
\Gamma\left(m-2+\frac{\epsilon}{2}\right)
\Gamma\left(1-\frac{\epsilon}{2}\right) \nl
&\times& \int [d \alpha]_m~{\cal X}_m^{(-1+\frac{\epsilon}{2})}\,.
\eqa
Since the constraint $\ell > 0$ in Eq. (\ref{eq:app1}) implies
$m-2> 0$, one could conclude that
\bqa
\label{eq:app8}
I^{(n;\,2(m-2))}_{m} = 0 + \ord(\epsilon)
\eqa
{\em if and only if} the integral over 
the Feynman parameters in the second line of  
(\ref{eq:app7}) does not contains poles in ${1}/{\epsilon}$.
One may wonder whether this can actually occur for there may be 
infrared and collinear divergences. A simple reasoning shows 
that this is never the case. In fact, when $\ell > 0$, the presence of
${\tilde q}^{2\ell}$ in the numerator always raises the powers of
the quadratic form ${\cal X}_m$ in the Feynman parameter integrals with
respect to the ``standard'' $\ell= 0$ loop functions, therefore 
forbidding the presence of collinear or soft divergences.
The same reasoning shows that
\bqa
\label{eq:app9}
I^{(n;\,2 (m-1+d))}_{m} = 0 + \ord(\epsilon)\,
~~~~{\rm when}~~d < 0\,.
\eqa
However, it is extremely instructive to compute Eq. (\ref{eq:app7})
directly.
For the most collinear and infrared divergent fully massless box diagram
in Fig. \ref{figure} one gets the finite expression
\bqa
\label{eq:app10}
\int [d \alpha]_3~\frac{1}{{\cal X}_3} = -\frac{1}{s+t}
\left[ \ln(s/t) \ln(-s/t) + {\rm Li_2}(1+t/s) + {\rm Li_2}(1+s/t)
\right]\,.
\eqa
Since all higher point loop functions can be rewritten as combinations
of box diagrams \cite{Tarasov:1996br,Binoth:1999sp}, this result 
explicitly proves Eq. (\ref{eq:app9}).
\begin{figure}
\bfig(300,130)
\SetScale{1}
\sof(-30,18)
\flin{40,0}{65,25} \lin{65,25}{65,70} \flin{40,95}{65,70}
\lin{65,25}{110,25}
\lin{65,70}{110,70}
\lin{110,25}{110,70}
\flin{110,70}{135,95}
\flin{110,25}{135,0}
\Text(37,0)[tr]{$p_3-p_1$}
\Text(37,95)[br]{$p_1$}
\Text(138,95)[bl]{$p_2$}
\Text(138,0)[tl]{$p_3-p_2$}
\Text(116,47)[l]{\footnotesize{0}}
\Text(60,47)[r]{\footnotesize{0}}
\Text(87,29)[b]{\footnotesize{0}}
\Text(87,74)[b]{$\leftarrow q $}
\Text(87,64)[t]{\footnotesize{0}}
\Text(185,65)[l]{$p_1^2\,=\, p_2^2\, =\, 0$}
\Text(185,45)[l]{$(p_3-p_1)^2\,=\, (p_3-p_2)^2\, =\, 0$}
\Text(185,25)[l]{$s = p_3^2\,,~~t = (p_1-p_2)^2 $}
\efig{Fig. \ref{figure}: {\em Fully massless box diagram.}}
\label{figure}
\end{figure}

We close this Appendix noting that, in any case,
a contribution ${\cal O}(1)$ can only develop for non-negative
powers of the quadratic form ${\cal X}_m$, so that the integrand
in the Feynman parameter integral is always polynomial.

\section{Three-point tensors: general case 
\label{appc}}
This Appendix extends Eq. (\ref{eq:3points}) to three-point tensor 
integrals of rank higher than 3.
We redefine
\bqa
d_\mu \equiv \frac{\beta}{\gamma} D_\mu ~~~{\rm and}~~~
L_\mu \equiv -\frac{1}{2\gamma}Q_\mu\,, 
\eqa
so that the first line of Eq. (\ref{eq:8}) simply 
reads $q_\mu= d_\mu+L_\mu$.
Then
\bqa
\label{eq:start}
q_{\mu_1} \cdots q_{\mu_i} = \sum_{k=1}^i 
\left\{L_{\mu_1} \cdots L_{\mu_{k-1}} \,d_{\mu_k} \,q_{\mu_{k+1}} 
\cdots q_{\mu_i}
\right\} + \prod_{k=1}^i L_{\mu_k}\,.
\eqa
The first term always contains reconstructed denominators 
and we do not elaborate it any further. 
Inserting the definition of $Q_\mu$ in Eq. (\ref{eq:8}) 
in the second term this can be written
\bqa
\label{eq:prod}
\prod_{k=1}^i L_{\mu_k}= \left( -\frac{1}{2\gamma}\right)^i
\sum_{k=0}^i (q \cdot\ell_3)^k (q \cdot\ell_4)^{(i-k)}\, 
S^{k}_{\mu_1 \cdots \mu_i}\,,
\eqa
where the tensor $S^{k}_{\mu_1 \cdots \mu_i}$ 
is defined to be the sum of all possible tensor products of 
$i$ 4-vectors $\ell_3$ and $\ell_4$, such that $\ell_4$ appears $k$ times.
For example,
\bqa
S^{2}_{\mu_1 \mu_2 \mu_3}= 
 \ell_{4\mu_1} \ell_{4\mu_2} \ell_{3\mu_3}
+\ell_{4\mu_1} \ell_{3\mu_2} \ell_{4\mu_3}
+\ell_{3\mu_1} \ell_{4\mu_2} \ell_{4\mu_3}\,.
\eqa
The two terms with $k= 0$ and $k= i$ in Eq. (\ref{eq:prod}) do not 
contribute to the integral, due to the Theorem in Eq. (\ref{eq:theorem1}). 
The product 
$(q \cdot \ell_3)(q \cdot \ell_4) =
\beta q^\lambda D_\lambda -\gamma q^2$
can be factorised out from all the remaining terms, 
giving as result
\bqa
\label{eq:prod2}
\prod_{k=1}^i L_{\mu_k} &=& \left(-\frac{1}{2\gamma}\right)^i\,
 [\beta q^\lambda D_\lambda -\gamma q^2]\,
q^{\alpha_1} \cdots q^{\alpha_{i-2}} \,
S_{[\alpha_1 \cdots \alpha_{i-2}]\,
      \mu_1 \cdots \mu_i}
\nl
S_{[\alpha_1 \cdots \alpha_{i-2}]\,
\mu_1 \cdots \mu_i}
&\equiv& \sum_{k=1}^{i-1}\, 
S^{i-k-1}_{\alpha_1 \cdots \alpha_{i-k-1}}\,
S^{0}_{\alpha_{i-k} \cdots \alpha_{i-2}}\,
S^{k}_{\mu_1 \cdots \mu_i}\,.
\eqa
Inserting Eq. (\ref{eq:prod2}) into Eq. (\ref{eq:start}), dividing
by the three denominators and integrating over $d^nq$ we obtain 
the desired relation
\bqa
I^{(n)}_{2;\,\mu_1 \cdots \mu_i} &=& \frac{\beta}{\gamma}\, 
\sum_{k=1}^i \left(-\frac{1}{2 \gamma }\right)^{k-1}\!\!
t^{\alpha_2}_{\mu_1} \cdots t^{\alpha_k}_{\mu_{k-1}}
\left \{ 
J^{(n)}_{2;\,\mu_k \alpha_2 \cdots \alpha_{k}\mu_{k+1}\cdots \mu_i}
\right \} 
+ \left(-\frac{1}{2\gamma}\right)^i
S^{[\alpha_1 \cdots \alpha_{i-2}]}_{\mu_1 \cdots \mu_i} \nl
&\times&
\left \{ 
\beta
J^{(n)\lambda}_{2;\,\lambda \alpha_1 \cdots \alpha_{i-2}}
-\gamma \left[m_0^2\,  I^{(n)}_{2;\,\alpha_1 \cdots \alpha_{i-2}} 
+I^{(n)}_{1;\,\alpha_1 \cdots \alpha_{i-2}}(0)
-I^{(n;\,2)}_{2;\,\alpha_1 \cdots \alpha_{i-2}} \right] 
\right \}\,,
\eqa
where $S^{[\alpha_1 \cdots \alpha_{i-2}]}_{\mu_1 \cdots \mu_i}$
is given in Eq. (\ref{eq:prod2}), but with the 
$\alpha$ indices lowered, and as in 
Eq. (\ref{eq:3points}) the tensors in the numerator of the 
$n$-dimensional integrals are 4-dimensional.

\end{document}